\documentclass[12pt]{article}

\usepackage{amssymb}
\usepackage{amsmath}

\newcommand{\pa}{\partial}

\newcommand{\myref}[1]{(\ref{#1})}

\newcommand{\de}{\delta}
\newcommand{\De}{\Delta}
\newcommand{\al}{\alpha}
\newcommand{\eps}{\varepsilon}
\newcommand{\la}{\lambda}
\newcommand{\La}{\Lambda}

\newcommand{\sig}{\sigma}

\renewcommand{\geq}{\geqslant}
\newcommand{\lan}{\langle}
\newcommand{\ran}{\rangle}

\newcommand{\demi}{\frac{1}{2}}

\newcommand{\sur}[2]{{\displaystyle\mathop{#1}_{#2}}}

\newcommand{\mcal}[1]{\mathcal{#1}}

\newlength{\somme}
\newlength{\sommep}
\newlength{\sommebis}
\newlength{\sommepbis}
\newcommand{\mC}{\mcal{C}}

\usepackage{anysize}
\usepackage{graphics}
\marginsize{2cm}{2cm}{2cm}{2cm}

\begin{document}
\title{Injected Power Fluctuations in 1D Dissipative Systems}
\author{Jean Farago$^1$\footnote{e-mail: farago@ics.u-strasbg.fr}\  and Estelle
  Pitard$^2$\footnote{e-mail: estelle.pitard@lcvn.univ-montp2.fr} \\
\null$^1$\textit{Institut Charles Sadron CNRS-UPR 22}\\
\textit{6 rue Boussingault BP 40016 F-67083 Strasbourg Cedex, France.}\\
\null$^2$\textit{Laboratoire des Verres, Colloides et Nanomat\'eriaux (CNRS-UMR 5587), CC69,}\\
\textit{ Universit\'e Montpellier 2, 34095 Montpellier Cedex 5, France.}}
\date{}
\maketitle
\abstract{Using fermionic techniques, we compute exactly the large
  deviation function (ldf) of the
  time-integrated injected power in several one-dimensional
  dissipative systems of classical spins. The dynamics are $T=0$ Glauber
  dynamics supplemented by an injection mechanism, which is taken as a
  Poissonian flipping of one particular spin. We discuss the physical content of
  the results, specifically the influence of the rate of the Poisson
  process on the properties of the ldf.}

\

{\bf Keywords:} Glauber dynamics, spin systems, large deviation
functions, free fermions, stochastic systems, fluctuation theorem.

\section{Introduction}
Fluctuations in nonequilibrium systems have attracted interest of
 physicists and mathematicians in recent years, due to fortuitous
 conjunctions of experiments\cite{fauveexp1,fauveexp2}, which aimed at measuring and understanding
 the fluctuations of global variables in hydrodynamic experiments, and
 theoretical works \cite{evanscohenmorris,gallavotticohen,kurchanrevue}, having stated new relations for the entropy
 production in some classes of nonequilibrium systems (the fluctuation
 theorems). This
 simultaneity  was actually the cause of some bold assertions
 surmising a wider application of the so-called fluctuation theorems
 to dissipative systems \cite{cili,feitosamenon}, whereas these relations were nothing but a
 manifestation of the time-reversal symmetry of the bulk
 dynamics. These conjectures were enforced by the apparent validity of
 the fluctuation relations in experiments and simulations, which
  yields incidentally
 a characteristic energy sometimes daringly termed ``nonequilibrium temperature''. Despite the
 fact that it has been convincingly proven that this apparent validity
 is only due to our inability of probing the large deviation functions
 of injected power in zones of negative injected power \cite{afmp,farago1,farago2,affm,pvbtw,vpbtw}, it is however
 interesting to show explicit examples of dissipative systems where the large deviation
 function (ldf) of injected power can be exactly computed, all the
 more so there exists very few such exact calculations on dissipative
 systems \cite{farago1,farago2,bertin,lrw,cornu}, for which the absence of detailed
 balance condition  makes the situation
 more difficult to handle.

 Moreover,
 for conservative systems in nonequilibrium stationary states, interesting relations on the related ldf of
 integrated current have been discovered recently \cite{bodineauderrida}, and it is 
 interesting to compare the classes of conservative and dissipative
 systems as regards the fluctuations of the global variable associated
 with the external excitation (the very cause of the
 nonequilibrium state).

\medskip

In this paper, we consider a 1D system of $2N$ ($N\rightarrow\infty$) classical spins on a ring, labelled from $-N$ to $N-1$, updated
according a $T=0$ Glauber dynamics, and supplemented by an injection
mechanism.
The $T=0$ Glauber dynamics is defined as follows : the probability for
a spin $s_j$ to flip between $t$ and $t+dt$ is given by
$dt[1-s_j(s_{j-1}+s_{j+1})/2]$, that is, $s_j$ flips with a rate 1 if
its neighbours are in different states, cannot flip if
$s_j=s_{j-1}=s_{j+1}$ and flips with a rate 2 if
$-s_j=s_{j-1}=s_{j+1}$. It is important to mention that these dynamics are
dissipative:  a flip either does not modify the energy of the system,
or lessens it. As a result, energy must be injected by an additional
mechanism of injection in order to drive the system into a non trivial stationary state.
We
will consider two different models of injection : for both models
(labelled  I and
II hereafter), only the zeroth spin $s_0$ (the ``external boundary'' of the
system) flips randomly as a Poisson process
with a parameter $\la$. The difference between the two models lies in
the ability of the zeroth spin,  besides the Poissonian flip, to spontaneously flip under the
Glauber rules (model II) or
not (model I); it is qualitatively more important than it
could appear, as in one case (model II) both halves of the system are dynamically
connected
in the vicinity of $s_0$ when in the other (model I) they are disconnected. In
the thermodynamic limit, Model I can be viewed as two independent
identical systems driven out of equilibrium by $s_0$ . 

For these systems, we computed exactly $f(p)$, the large
deviation function of the injected energy, a major observable
associated with the fluctuations of the energy flux in stationary
systems. It is defined as 
\begin{align}
  f(p)=\lim_{\tau\rightarrow\infty}\tau^{-1}{\log \text{Prob}(\text{(energy
  injected between 0 and $\tau$)}/\tau=p)}
\end{align}
As the dynamics are stochastic, no explicit formula can be put forward
for $\Pi$ the energy injected in the system by $\la$ from $t=0$ to
$t=\tau$, but fortunately such expressions are not needed for our purpose.
In section \ref{calcul}, we expose the dynamics of the models in
greater detail, as well as the computation and the exact results for $g(\al)$, the ldf
of the characteristic function. In 
section \ref{ldff} we numerically solve the inverse Legendre transforms and get
the ldfs for the injected energy for the two models considered here. We discuss their physical
properties, in particular their variations with respect to the
change of the injection rate.

\section{Fermionic approach to the injected power}\label{calcul}

 It is useful to describe spin systems in the
dual representation of the domain walls : between the site 0 and 1 is
located the locus of a possible domain wall labelled $0$, and so forth. The state
of the system is thus characterized by
$\mcal{C}=(n_{-N},\ldots,n_{-1},n_0,n_1,n_2,\ldots,n_{N-1})$, where the
$n_i$ are either 0 (no domain wall) or 1. There is $2^{2N}$ possible
states in this representation, in contrast with the spin
representation where a degeneracy $s\leftrightarrow -s$ doubles this number. The dynamical equation for the probability is given by
\begin{align}
  \pa_tP(\mcal{C})=\la [P(\mcal{C}_0)-P(\mcal{C})]+\sum_{j}
  [P(\mcal{C}_j)w(\mcal{C}_j\rightarrow \mcal{C})-P(\mcal{C})w(\mcal{C}\rightarrow \mcal{C}_j)]
\end{align}
where $\mcal{C}_j$ holds for the state $\mcal{C}$ whose domain
wall variables $n_j$ and $n_{j-1}$ have been changed (according to
$n\rightarrow 1-n$). The $T=0$
Glauber dynamics corresponds to
\begin{align}
  w(\mcal{C}_j\rightarrow \mcal{C})&=2-n_j-n_{j-1}\\
  w(\mcal{C}\rightarrow \mcal{C}_j)&=n_j+n_{j-1}
\end{align}
where $n_j$ and $n_{j-1}$ are the variables associated with the state
$\mcal{C}$ (we use this convention hereafter). 
Note that models I and II differ in the preceding equations in the way
the summation over $j$ is carried out: either it is not restricted
(model II), or index $j=0$
is implicitely removed (model I).

\medskip

We consider that each domain wall contributes as  a
quantum of energy $1$ to the global energy of the system.
We are interested in the energy $\Pi$ injected into the system up to time $t$ by the
poissonian injection. Following \cite{derridalebowitz}, the route to this time integrated observable
begins with the consideration of the joint probability
$P(\mcal{C},\Pi,t)$, the probability for the system to be in the state
$\mcal{C}$ at time $t$ having received the energy $\Pi$ from the
injection. The dynamical equation for this quantity is readily
\begin{multline}
  \pa_tP(\mcal{C},\Pi)= \la
  \{P(\mcal{C}_0,\Pi-2)n_0n_{-1}+P(\mcal{C}_0,\Pi+2)(1-n_0)(1-n_{-1})\\
+P(\mcal{C}_0,\Pi)[(1-n_0)n_{-1}+(1-n_{-1})n_0]  - P(\mcal{C},\Pi)\}
+\sum_{j}
  [P(\mcal{C}_j,\Pi)w(\mcal{C}_j\rightarrow \mcal{C})-P(\mcal{C},\Pi)w(\mcal{C}\rightarrow \mcal{C}_j)]
\end{multline}
We define next the  generating function of $\Pi$
as
\begin{align}
  F(\mcal{C})=\sum_{\Pi=-\infty}^{\infty}e^{\al\Pi}P(\mcal{C},\Pi)
\end{align}
This quantity, summed up over the states, yields the generating
function $\lan \exp(\al \Pi)\ran$ from which one derives its ldf
$g(\al)$:
\begin{align}
  \lan e^{\al \Pi}\ran\sur{\simeq}{t\rightarrow\infty}e^{t g(\al)}
\end{align}
This ldf $g(\al)$ is closely related to $f(p)$, the ldf of the probability
density function of $\Pi$, as they are Legendre transform of each
other \cite{farago1} :
\begin{align}
  \text{Prob}(\Pi/t=p)&\sur{\propto}{t\rightarrow\infty}\exp(tf(p))\\
f(p)&=\min_{\al}\left(g(\al)-\al p\right)
\end{align}

Let us write the dynamical equation for $F$ :
\begin{multline}
  \pa_t F(\mcal{C})=\\
\la\left[
  e^{2\al}F(\mcal{C}_0)n_0n_{-1}+e^{-2\al}F(\mcal{C}_0)(1-n_0)(1-n_{-1})+F(\mcal{C}_0)[(1-n_0)n_{-1}+(1-n_{-1})n_0]-F(\mcal{C})\right]\\
+\sum_j [F(\mcal{C}_j)(2-n_j-n_{j-1})-F(\mcal{C})(n_j+n_{j-1})]
\end{multline}
The function $g(\al)$ is closely related to the dynamical matrix at
work in
the rhs of the preceding equation, since it is in general its largest
eigenvalue. Thus, succeeding in computing $g(\al)$ is not as
complicated  as knowing the whole dynamics (eigenvalues \textit{and} eigenvectors), but nevertheless it
remains a difficult challenge, as the whole spectrum of the matrix
 has to be
known.

Fortunately, our problem belongs to the category of the
``free-fermions'' problems, for which a diagonalization of the
dynamics into independent ``modes'' can be achieved. To do so, we
construct the state vector as
\begin{align}
  |\phi\ran&=\sum_\mcal{C}F(\mcal{C})|\mC\ran
\end{align}
where $|\mC\ran$ is a vector in the $2^{2N+1}$ dimensional tensorial product of the space of
states of domain walls (each of dimension 2). The dynamical evolution
of $|\phi\ran$ yields expressions like
$  \sum_\mcal{C}F(\mC_0)n_0n_{-1}|\mC\ran$
which can be represented as
\begin{align}
    \sum_\mcal{C}F(\mC_0)n_0n_{-1}|\mC\ran&=\sum_\mC
    F(\mC_0)\widehat{n}_0\widehat{n}_{-1}|\mC\ran\\
&=\widehat{n}_0\widehat{n}_{-1}\sum_\mC F(\mC)|\mC_0\ran\\
&=\widehat{n}_0\widehat{n}_{-1}\sig^x_0\sig^x_{-1}|\phi\ran=s^-_0s_{-1}^-|\phi\ran
\end{align}
where
\begin{align}
\widehat{n}=  \left(\begin{array}{cc}0&0\\0&1\end{array}\right), \ \ 
\sigma^x=  \left(\begin{array}{cc}0& 1\\1& 0\end{array}\right),\ \ s^-=  \left(\begin{array}{cc}0& 0\\1& 0\end{array}\right)
\end{align}
(we will have use also of $s^+=\null^t(s^-)$ ; note that $s^-s^+=\widehat{n}$). Note that for the time
being, operators with different particle indices commute. Similar
computations on other terms allows to write the dynamical equation for $|\phi\ran$ as
\begin{align}
\pa_t  |\phi\ran&=H|\phi\ran\ \ \text{where}\\
H&=\la[e^{2\al}s_0^-s_{-1}^-+e^{-2\al}s_0^+s_{-1}^++s_0^+s_{-1}^-+s_0^-s_{-1}^+-1]\nonumber\\
&\ \ \ \ +\sum_j[2s_{j-1}^+s_{j}^++s_{j-1}^+s_{j}^-+s_{j-1}^-s_{j}^+-s_{j}^-s_{j}^+-s_{j-1}^-s_{j-1}^+]
\end{align}
The fermionization of this ``Hamiltonian'' proceeds using the
Jordan-Wigner transformation
\begin{align}
  c_{-N}&\sur{=}{\text{def}}s_{-N}^+\\
  c_{-N}^\dag&\sur{=}{\text{def}}s_{-N}^-\\
  c_j&\sur{=}{\text{def}}s_j^+\sig_{-N}^z\sig_{-N+1}^z\ldots\sig_{j-1}^z\\
  c_j^\dag&\sur{=}{\text{def}}s_j^-\sig_{-N}^z\sig_{-N+1}^z\ldots\sig_{j-1}^z
\end{align}
where 
\begin{align}
\sig^z=  \left(
  \begin{array}{lr}
    1&0\\0&-1
  \end{array}\right)
\end{align}
is the third Pauli matrix.
A major change provided by this transformation is that now operators
with different indices do no longer commute. Instead, it can be
verified that for all $i$ and $j$
\begin{align}
  \{c_i,c_j\}&=0\\
  \{c_i^\dag,c_j^\dag\}&=0\\
  \{c_i^\dag,c_j\}&=\de_{i,j}
\end{align}
where the bracket is defined by $\{a,b\}=ab+ba$. These relations,
characteristic of fermionic operators, yields a useful rewriting of
the dynamical operator
\begin{align}
  H=\la[e^{2\al} c_{-1}^\dag c_0^\dag+e^{-2\al} c_0c_{-1}+c_0^\dag c_{-1}+c_{-1}^\dag
  c_0-1]+\sum_j [2c_jc_{j-1}+c_j^\dag c_{j-1}+c_{j-1}^\dag
  c_j-c_j^\dag c_j-c_{j-1}^\dag c_{j-1}]
\end{align}
The change of variables $\tilde{c}^\dag=e^\al
c^\dag,\tilde{c}=e^{-\al} c$ is compatible with the fermionic
structure and allows the rewriting of the preceding equation (we omit
the tildas in the following) :
\begin{align}
  H=\la[c_{-1}^\dag c_0^\dag+c_0c_{-1}+c_0^\dag c_{-1}+c_{-1}^\dag
  c_0-1]+\sum_j [2e^{2\al}c_jc_{j-1}+c_j^\dag c_{j-1}+c_{j-1}^\dag
  c_j-c_j^\dag c_j-c_{j-1}^\dag c_{j-1}]
\end{align}
In order to achieve the diagonalization, we follow the route described
in the Appendix. The Hamiltonian can be written as
\begin{align}
  H&=   \sum_q\La_q\left(\xi^\dag_q\xi_q-\demi\right)-2N+1-\eps-\la
\end{align}
where $\eps$ is defined as:
\begin{align}
\eps&=\left\{
\begin{array}{ll}
  0 & \text{(model I)}\\
  1 & \text{(model II)}
\end{array}\right.
\end{align}
The largest eigenvalue of $H$ is thus given by
\begin{align}
  g(\al)&=\demi\sum_q|\La_q|-2N+1-\eps-\la
\end{align}
provided all the $\La_q$ are real. If one takes into account possible
complex eigenvalues, $g(\al)$ will be rather defined by
\begin{align}\label{formalg}
  g(\al)=\demi\sum_q|\text{Re}(\La_q)|-2N+1-\eps-\la
\end{align}
As explained in the appendix,
$\{\La_q,-\La_q\}_q$ is the spectrum of the matrix
\begin{align}
  M_0=\left(
  \begin{array}{lr}
    A & B\\ D & -A   
  \end{array}\right)
\end{align}
with the notations of the appendix; moreover we can compute its
characteristic polynomial from \myref{app47}, taking advantage of
 $B$  almost empty. We define $E(\mu)=(A+\mu)^{-1}D(A-\mu)^{-1}$
and \myref{app47} shows that we have to compute $\det(1+BE)$; the matrix $(1+BE)$ is quite
simple, since $B=\la(M_{-1,0}-M_{0,-1})$, where $\{M_{ij}\}_{ij}$ is the
canonical basis of the $2N\times2N$ matrices. As a result $BE$ is zero
except on the lines $-1$ and $0$, and one has
\begin{align}
\det(1+BE)&=(1-\la E_{-1,0})(1+\la E_{0,-1})+\la^2 E_{0,0}E_{-1,-1}\\
&=(1+\la E_{0,-1}(-\mu))(1+\la E_{0,-1}(\mu))+\la^2 E_{0,0}(\mu)E_{-1,-1}(\mu)
\end{align}
where we exploited the fact that $E^T(\mu)=-E(-\mu)$. Note in passing
that the symmetry of this expression with respect to
$\mu\rightarrow-\mu$, demonstrated with general arguments in the
appendix, is blatant here, as the $E_{jj}$ are antisymmetric functions
of $\mu$. 

A further simplification is provided by the fact that
the physical system has itself a symmetry $i\leftrightarrow -i-1$ (expressed in the matrices $A$, $B$ and $D$ as $X_{i,j}=X_{-j-1,-i-1}$)
which gives $E_{00}(\mu)=-E_{-1,-1}(\mu)$ and
$E_{0,-1}(\mu)=E_{0,-1}(-\mu)$, whence
\begin{align}
  \det(1+BE)&=[1+\la E_{0,-1}(\mu)+\la E_{0,0}(\mu)][1+\la
  E_{0,-1}(\mu)-\la E_{0,0}(\mu)]
\end{align}

The next step is to compute the elements $(0,j)$ and $(-1,j)$ of the
matrix $(\mu \text{Id}-A)^{-1}$, or equivalently the associated
cofactors of $\mu\text{Id}-A$, termed $\text{Cof}(i,j)$. Precisely, we need the cofactors
$\text{Cof}(0,j)$, and the others are obtained using
$\text{Cof}(i,j)=\text{Cof}(j,i)$ and $\text{Cof}(i,j)=\text{Cof}(-i-1,-j-1)$.
We get :
\begin{align}
  [(\mu-A)^{-1}]_{0,j}=\frac{\text{Cof}(0,j)}{\det(\mu-A)}\sur{\sim}{N\rightarrow\infty}\left\{
  \begin{array}{lr}
\displaystyle\frac{[\la+\eps](-1)^{-j}x^{j+1}}{(x+\eps-1)^2-(\la+\eps)^2} & \text{if $j<0$} \\
\displaystyle\frac{[x+\eps-1](-1)^jx^{-j}}{(x+\eps-1)^2-(\la+\eps)^2} & \text{if $j\geq0$}
 \end{array}\right.
\end{align}
where
\begin{align}
x&=\frac{\mu+2\pm\sqrt{\mu^2+4\mu}}{2}\  \
\text{($\pm$ such that $|x|$ be maximum)}\label{eqx}
\end{align}
After cumbersome computations, we get, for $N\rightarrow\infty$ :
\begin{align}\label{chiM0}
  \chi_{M_0}(\mu)&= G_+(\mu)G_-(\mu)\frac{(x_+x_-)^{2N}}{\mu^4-16\mu^2}\\
 G_+(\mu)&= 
  (x_+-1+2\eps+\la)(x_--1-\la)+\la\left[\la+2\eps
 e^{2\al}+2e^{2\al}\frac{x_--x_+}{x_+x_--1}\right]\\
G_-(\mu)&=  (x_--1+2\eps+\la)(x_+-1-\la)+\la\left[\la+2\eps
 e^{2\al}+2e^{2\al}\frac{x_+-x_-}{x_+x_--1}\right]\\
x_+&=\frac{\mu+2+\sqrt{\mu^2+4\mu}}{2}\ \ \text{for Re$(\mu)>-2$}\\
x_-&=\left\{
\begin{array}{ll}
  \frac{-\mu+2+\sqrt{\mu^2-4\mu}}{2} & \text{for Re$(\mu)\in[0,2]$}
  \\
  \frac{-\mu+2-\sqrt{\mu^2-4\mu}}{2} & \text{for Re$(\mu)>2$}
\end{array}
\right.
\end{align}

Finally, one has to combine this last result with
\myref{formalg}. This is done via the formula
\begin{align}
g(\al)&=\frac{1}{4i\pi}\oint d\mu \ \left[\mu\frac{\chi_{M_0}'(\mu)}{\chi_{M_0}(\mu)}\right] -2N+1-\eps-\la
\end{align}
where the contour for the complex integration is a big (big enough to
enclose all singularities of the meromorphic function)
half-circle leant on the imaginary axis, with its belly in the
$\text{Re}(\mu)>0$ region, followed  counterclockwise. The logarithmic derivative
of $\chi_{M_0}$ yields different additive contributions from formula
\myref{chiM0}. Let us consider first the contribution
$(x_+x_-)^{2N}$. The integral
\begin{align}
  I=\frac{2N}{4i\pi}\oint d\mu\left[ \mu\frac{x_+'}{x_+}+\mu\frac{x_-'}{x_-}\right]\end{align}
can be exactly computed, as the analytical singularities in the right
half plane of $x_+$ and
$x_-$ are located only on the real segments $[-4,0]$ and $[0,4]$ respectively; as a result, we
can compute this integral with a contour encircling $[0,4]$ and
sticking to it. Yet some care must
be taken however, for the actual expression of $x_-$ is not uniform on
the whole half plane, and notably on the real axis: whereas $x_+$ is
everywhere given on $\text{Re}(\mu)>0$ by $[\mu+2+\sqrt{\mu^2+4\mu}]/2$, $x_-$
is given by $[-\mu+2+\sqrt{\mu^2-4\mu}]/2$ for $\text{Re}(\mu)\in[0,2]$
and by $[-\mu+2-\sqrt{\mu^2-4\mu}]/2$ for $\text{Re}(\mu)>2$.
With all these precautions we get the simple result $I=2N$.
This extensive term vanishes with the corresponding term coming from
$\text{Tr}(A)$, which is coherent with the injection properties being
asymptotically independent of the size of the system (in our 1D
model). 

We can also compute easily the contribution of the term
$(\mu^4-16\mu^2)^{-1}$. The associated integral yields $-2$, thence we
can recast the expression for $g(\al)$ into
\begin{align}\label{theformula}
  g(\al)&=\frac{1}{4i\pi}\oint d\mu \ \mu\left[\frac{G'_+(\mu)}{G_+(\mu)}+\frac{G'_-(\mu)}{G_-(\mu)}\right]-1-\eps-\la
\end{align}

\subsection{The case $\al=0$}

It is possible to verify partially the correctness of the
formula. Indeed, when $\al=0$, we know the value of $g(\al)$: $g(0)=0$, that maximum eigenvalue being associated with the
stationary distribution (encoded in the corresponding
eigenvector). Thus, we must verify this expected result, which holds
whatever the values of $\la$ and $\eps$.

To do that, we rewrite the expressions for $G_+$ and $G_-$ in a more
transparent form. Using the relations $x_\pm^2-(\pm\mu+2)x_\pm+1=0$,
one can write
\begin{align}
  G_+(\mu)=(x_--1)\left[(x_+-1)(1+2\la/\mu)+2\eps\right]+2\la\frac{e^{2\al}-1}{x_+x_--1}[\eps(x_+x_--1)+x_--x_+]
\end{align}
$G_-$ is easily obtained by switching $x_+$ and $x_-$, \textit{and} making
$\mu\rightarrow-\mu$:
\begin{align}
  G_-(\mu)=(x_+-1)\left[(x_--1)(1-2\la/\mu)+2\eps\right]+2\la\frac{e^{2\al}-1}{x_+x_--1}[\eps(x_+x_--1)+x_+-x_-]
\end{align}

\subsubsection{$\eps=0,\al=0$}
This case is the simplest, since $G_\pm=(x_--1)(x_+-1)(1\pm2\la/\mu)$
and the logarithmic derivative yields six  contributions. The
two contributions coming from $(x_+-1)$ and the contribution coming
from $(1+2\la/\mu)$ yield zero, as no branch cut/poles are associated with
these functions in the half plane Re$(\mu)>0$. Thus, from
\myref{theformula},
\begin{align}
  g(0)&=\frac{1}{4i\pi}\oint d\mu \mu
  \left(\frac{2x_-'}{x_--1}+\frac{1}{\mu-2\la}\right)-1-\la\\
&=\frac{1}{4i\pi}\oint d\mu \mu
  \frac{2x_-'}{x_--1}-1
\end{align}
The contour of integration can be stuck to the branch cut $[0,4]$ (that
is $[0\rightarrow 4]-i0$ followed by $[4\rightarrow 0]+i0$. From $0-i0$ to
$2-i0$,
$x_-(\mu=y-i0)=[-\mu+2+\sqrt{\mu^2-4\mu}]/2=[-y+2+i\sqrt{4y-y^2}]/2$;
from $2-i0$ to $4-i0$, the expression for $x_-(\mu)$ is changed, but
the limit for $\mu$ sticking below the branch cut is such that one has
still $x_-(\mu=y-i0)=[-y+2+i\sqrt{4y-y^2}]/2$. The expressions above
the branch cut are just the complex conjugates. As a result $x_-$ covers
counterclockwise
the set of complexes of modulus $1$, from $z=1$ ($\mu=0-i0$) to $z=-1$
($\mu=4-i0$) with positive imaginary parts, and from $z=-1$
($\mu=4+i0$) back to $z=1$ with negative imaginary parts. As a result,
we get the expected value:
\begin{align}
  g(0)
&=\frac{1}{2i\pi}\oint_{|z|=1} dz \frac{2-z-z^{-1}}{z-1}-1=0
\end{align}

\subsubsection{$\eps=1,\al=0$}
The computation in that case is slightly different, in particular as
regards the localization of the singularities. The terms $(x_+-1)$ and
$(x_++1+2\la\mu^{-1}(x_+-1))$ have no singularities in the half plane
under consideration (the singularity $\mu=0$ is avoided). The term $(x_--1)$ yields a factor $1/2$, as
already computed. The term 
\begin{align}
x_-+1-2\la\mu^{-1}(x_--1)=-\frac{x_--1}{x_-}(x_-^2+2\la x_--1)
\end{align}
 is the
richest, as it has not only a branch cut, but also displays an isolated pole for
$\mu=2[1+\sqrt{\la^2+1}]$. Thus,
\begin{align}
  g(0)&=-\frac{1}{4i\pi}\oint d\mu \mu
  \frac{x_-'}{x_-}+\frac{1}{2i\pi}\oint d\mu \mu
  \frac{x_-'}{x_--1}-\frac{1}{2i\pi}\oint_{|z|=1} \frac{dz}{z}\
  \frac{(z-1)^2(z+\la)}{z^2+2\la z-1}-1+\sqrt{\la^2+1}-\la\\
&=-\frac{1}{2i\pi}\oint_{|z|=1} \frac{dz}{z}\
  \frac{(z-1)^2(z+\la)}{z^2+2\la z-1}-1+\sqrt{\la^2+1}-\la=0
\end{align}

\subsection{The general case ($\al\neq 0$)}

The preceding particular cases show clearly that the analysis of the
singularities becomes very complicated in the general case, as the new
term proportional to $e^{2\al}-1$ mixes  $x_+$ and $x_-$. It is
thus more appropriate to switch to an integral representation of the
result. To achieve this in a simple way, we make use of the preceding
verifications. We define
\begin{align}\label{Iplus}
  I_+(\mu)=1+\frac{2\la(e^{2\al}-1)}{x_-x_++1}\times\frac{1+\eps(x_-x_+-1)/(x_--x_+)}{\mu/2+\la+\eps\mu/(x_+-1)}
\end{align}
and similarly $I_-$ using the rules $x_-\leftrightarrow x_+$ and
$\mu\leftrightarrow -\mu$. Using the identity
\begin{align}
  \frac{(x_--x_+)(x_-x_++1)}{(x_+-1)(x_--1)(x_-x_+-1)}=2/\mu
\end{align}
one shows that
\begin{align}\label{superultimate}
  g(\al)=\frac{1}{4i\pi}\oint d\mu \ \ \mu\left[\frac{I_+'}{I_+}+\frac{I_-'}{I_-}\right]
\end{align}

\subsubsection{Model I ($\eps=0$)}
The simplest expression is  provided by the system I ($\eps=0$) since
in that case:
\begin{align}
  I_\pm=1+\frac{2\la(e^{2\al}-1)}{x_-x_++1}\times\frac{1}{\pm\mu/2+\la}
\end{align}
The integrand of \myref{superultimate} is of order $\mu^{-3}$, so
 the half circle of integration can be made infinite, the only
non vanishing contribution being that of the vertical axis
Re$(\mu)=0$. It is easy to verify that on this axis $x_+=x_-^*$ and
$x_-(-iy)=x_-(iy)^*$. As a result, one has
\begin{align}
  g(\al)&=\frac{1}{4\pi}\int_{\infty}^{-\infty}dy\ y\
  \left[\frac{1}{I_+}\frac{dI_+}{dy}+\frac{1}{I_-}\frac{dI_-}{dy}\right]\\
&=\frac{1}{4\pi}\int_{-\infty}^\infty dy \log(|I_+|^2)=\frac{1}{2\pi}\int_0^\infty dy \log(|I_+|^2)
\end{align}
The explicit expression for $|I_+|^2$ is:
\begin{align}
  |I_+(iy)|^2&=1+\frac{4\la^2(e^{2\al}-1)(x_-x_++e^{2\al})}{(x_-x_++1)^2(\la^2+y^2/4)}
\end{align}
and
\begin{align}
  x_-x_+&=|x_+|^2=\frac{1}{4}\left|iy+2+\sqrt{-y^2+4iy}\right|^2\\
&\sur{=}{y\geq0}\frac{1}{4}\left[\left(2+\frac{\sqrt{y}}{\sqrt{2}}\sqrt{-y+\sqrt{y^2+16}}\right)^2+\left(y+\frac{\sqrt{y}}{\sqrt{2}}\sqrt{y+\sqrt{y^2+16}}\right)^2\right]
\end{align}
To summarize, we can write the result as
\begin{align}\label{theresult}
  &{g(\al)=\frac{2}{\pi}\int_0^\infty du\
  \log\left(1+\frac{\la^2(e^{2\al}-1)(\psi(u)+e^{2\al})}{(\psi(u)+1)^2(\la^2/4+u^2)}\right)}\\
&\psi(u)=\left(1+\sqrt{2}\sqrt{-u^2+\sqrt{u^4+u^2}}\right)^2+\left(2u+\sqrt{2}\sqrt{u^2+\sqrt{u^4+u^2}}\right)^2\label{psi}
\end{align}
A useful relation for $\psi(u)$ can be deduced: from 
\begin{align}
  \frac{1}{x_-x_++1}=\frac{1+x_+/x_-}{(1+x_-x_+)(1+x_+/x_-)}=\frac{1}{4}\left[\frac{1}{x_+}+\frac{1}{x_-}\right]
\end{align}
one gets
\begin{align}
  \frac{1}{\psi(u)+1}=\demi-\frac{1}{\sqrt{2}}\sqrt{-u^2+\sqrt{u^4+u^2}}
\end{align}
This suggests the change of variables $u=1/\sinh \tau$. This change
gives the alternate expression:
\begin{align}
{  g(\al)=\frac{2}{\pi}\int_0^\infty d\tau\ \frac{\cosh\tau}{\sinh^2\tau}\log\left(1+\frac{\la^2(e^{2\al}-1)(\coth^2\frac{\tau}{4}+e^{2\al})}{(\coth^2\frac{\tau}{4}+1)^2(\frac{\la^2}{4}\sinh^2\tau+1)}\sinh^2\tau\right)}
\end{align}
A partial verification of the result can be done, via the computation
of $g'(0)$. This quantity equals the mean injected power in the
stationary state, which can be computed directly by other means
\cite{farago3}. We have readily from \myref{theresult} :
\begin{align}
  g'(0)&=\frac{4\la^2}{\pi}\int_0^\infty
  \frac{du}{(\psi(u)+1)(\la^2/4+u^2)}\\
&=2\la(1+\la-\sqrt{\la^2+2\la})\label{meanvalueI}
\end{align}
which is the expected result. Note that it is not surprising to find exactly the
same result as in \cite{farago3} despite the fact that here the system
is duplicated with respect to the one studied in \cite{farago3}. There, the energy
associated with a domain wall was twice the value adopted here ($1$
per domain wall).

\subsubsection{Model II ($\eps=1$)}

This model yields an expression slightly more complicated for
$g(\al)$, due to an involved expression for $I_+$ (formula
\myref{Iplus}). Some algebraic manipulations allow only a partial
simplification :
\begin{align}
  I_+(\mu)=1+\frac{2\la(e^{2\al}-1)}{\mu\psi(\psi+1)}\times\frac{2\mu\psi-(\psi-1)^2}{\sqrt{\mu^2+4\mu}+2\la}
\end{align}
where $\psi=x_+x_-$. This yields the result
\begin{align}
  g(\al)=\frac{2}{\pi}\int_0^\infty du \log\left|1+\frac{\la(e^{2\al}-1)}{\psi+1}\times\frac{1+i(\psi-1)^2/8u\psi}{\la/2+\sqrt{-u^2+iu}}\right|^2
\end{align}
with $\psi=\psi(u)$  given by \myref{psi}. This seemingly irreducible
complexity we could have expected, since in contrast with  model I,
the mean injected power is here not computable by elementary
operations (i.e. the dynamical equations for the correlators
$\lan\sig_0\sig_i\ran$ are not closed).

\section{The large deviation functions}\label{ldff}

In this section, we discuss the properties of the ldf $f(p)$
associated with the integrated injected energy :
\begin{align}
  \text{Prob}(\Pi/t=p)&\sur{\propto}{t\rightarrow\infty}\exp(tf(p))\\
f(p)&=\min_{\al}\left(g(\al)-\al p\right)
\end{align}

\begin{figure}
  \centerline{\resizebox{12cm}{!}{\includegraphics{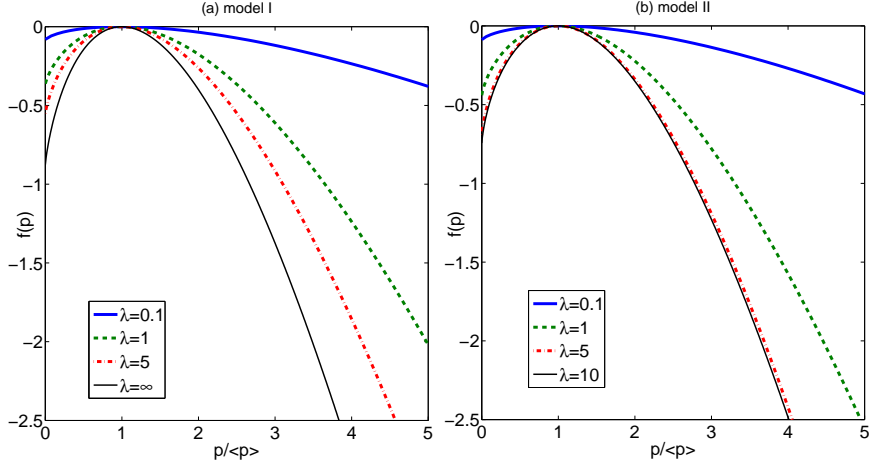}}}
\caption{Typical shapes of the large deviation functions of model I
  (left) and II (right),
  for various values of $\la$. The abscissae only are rescaled. For
  model II, $\la=\infty$ is not plotted due to the particular analytical
  behaviour of $f$ in that case (see text); however we believe $\la=10$ to be close
  to the asymptotic case.}\label{typsha}
\end{figure}

\begin{figure}
  \centerline{\resizebox{6cm}{!}{\includegraphics{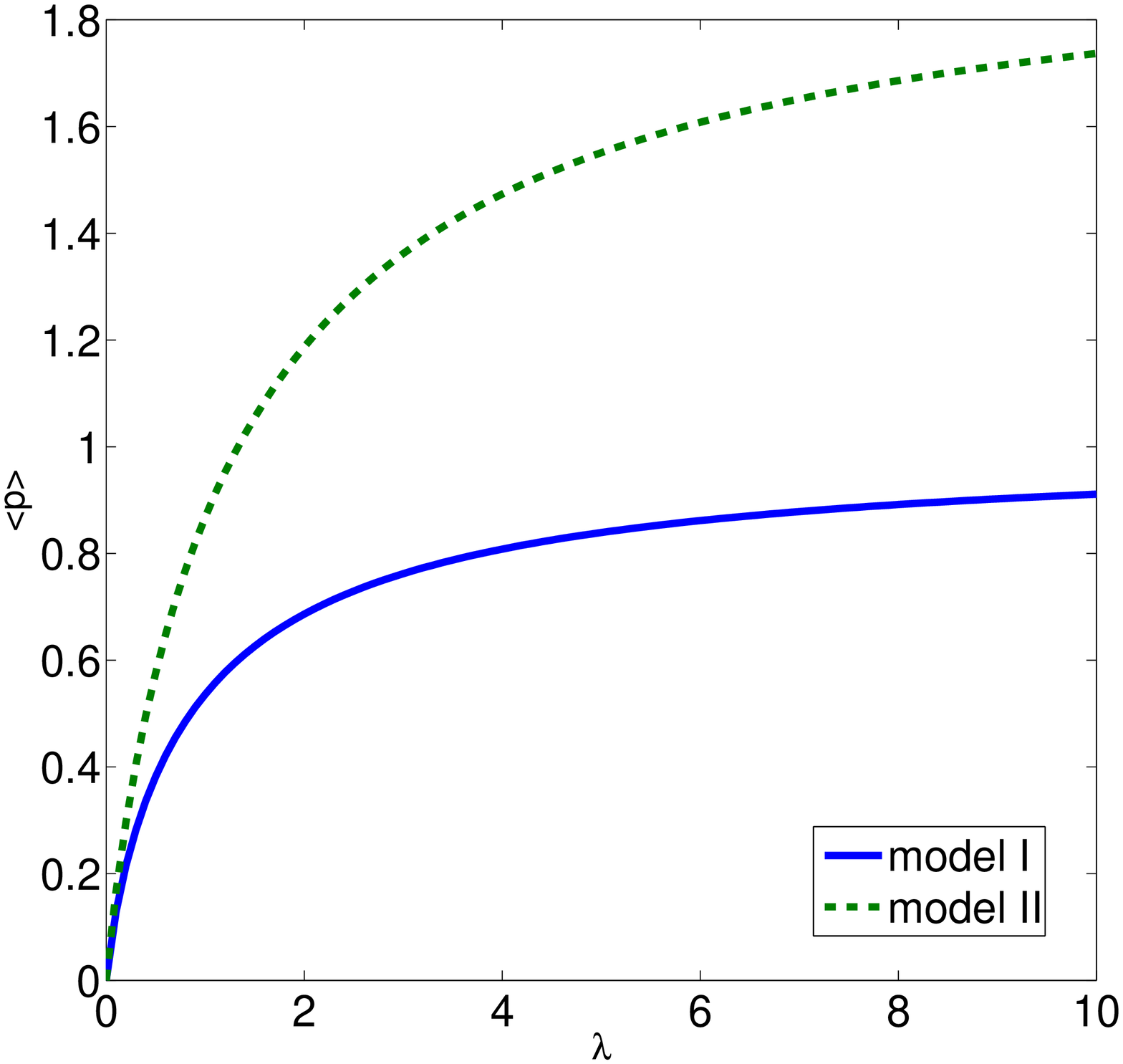}}}
\caption{Mean injected power as a function of $\la$ for the two models
considered.}\label{meaninjectedpower}
\end{figure}

The typical shapes of the large deviation functions are plotted on
figure \ref{typsha}. Their abscissae are rescaled to align their
maximum on 1. The missing information on $\lan p\ran$ is plotted on
figure \ref{meaninjectedpower} for both models (obtained via $\lan p\ran=g'(\al=0)$).
 The mean injected power behaves
 monotonously in both models
(fig. \ref{meaninjectedpower}), with but
a maximal asymptotic value twice larger (2) for model II.
We verified the consistency of this result by a direct simulation of
the model II (results not shown; for  model I, we have the exact
formula $\lan p\ran=2\la(\la+1-\sqrt{\la^2+2\la})$). That  model II induces an injection
larger than model I is obvious: the spin $s_0$ can switch back
spontaneously and swiftly just after the creation of two domain walls,
putting the system back in a state ready to accept anew a positive
injection of energy; in  model I, the domain walls have to move
away diffusively
before such a state can be reached. It is interesting to note that for
model II $\lim_{\la\rightarrow\infty}g'(0)$ is not given by the
simple limit taken inside the integral (which would have given $\lan
p\ran=1$). We postpone the discussion on this peculiarity to the end of the section.

\medskip

The ldfs for both models have the interesting property that they have a finite
limit when $p\rightarrow 0$: arbitrary small integrated injected energies ``are
not so rare''; this feature would a priori favour the presence of a
negative tail in the case where one averages over different initial conditions (as described in
\cite{farago1}); whether it is or not the case here  is an issue beyond the scope of this paper.

\begin{figure}[h]
  \centerline{\resizebox{6cm}{!}{\includegraphics{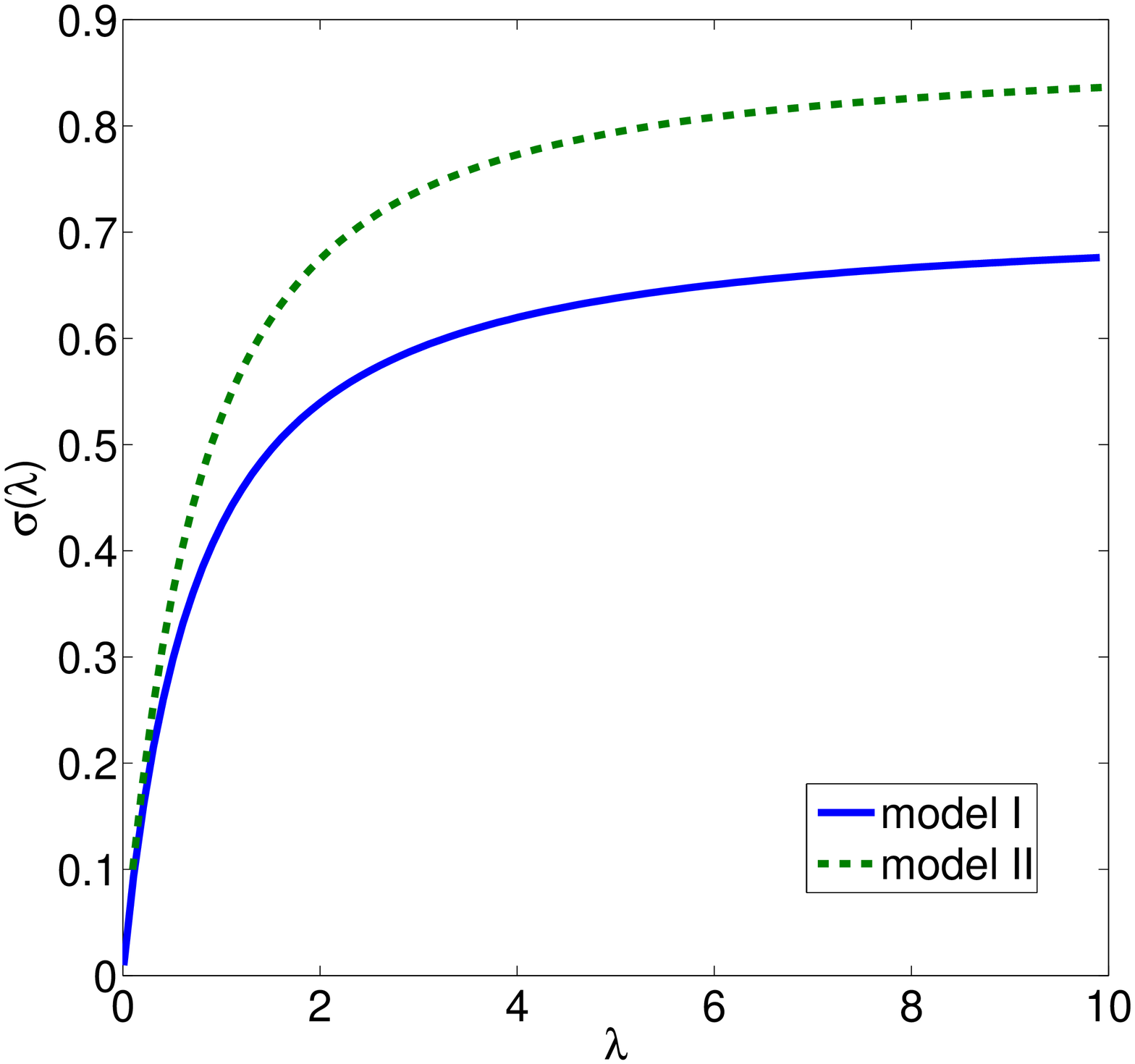}}}
\caption{Parameter $\sig(\la)=|f''(\lan p\ran)|\times\lan p\ran^2$ as a function of $\la$.}\label{relfluc}
\end{figure}

The ldfs are apparently close to a parabola. This is of course not
true, and the deviation from the parabola is a rather important
feature to look at. To this end, one defines $\sig(\lambda)=|f''(\lan
p\ran)|\times\lan p\ran^2$, which inverse is  a measure of the
relative fluctuations of $\Pi$ (up to a $1/\sqrt{\tau}$ factor). We see from figure \ref{relfluc} that
$\sigma(\lambda)$  increases gently with $\la$ and reaches a constant value.
This is consistent with the shrinking of $f$ with
increasing $\la$ to be seen on figure \ref{typsha}.

\begin{figure}[h]
  \centerline{\resizebox{6cm}{!}{\includegraphics{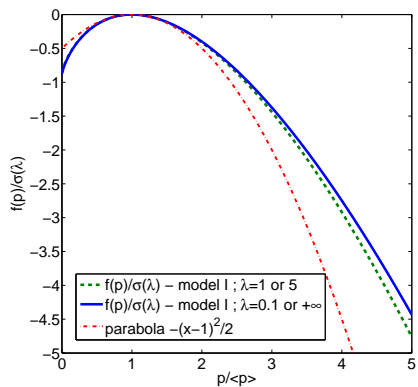}}}
\caption{Rescaled large deviation functions for the model I. Note that
  the curves for $\la=0.1$ and $\la=\infty$ are almost identical, and
  this also true for $\la=1$ and $\la=5$. This is related to the non-monotonous
  behaviour of the parameter $\chi(\lambda)$, see text.}\label{res_I_rescaled_xy}
\end{figure}

\begin{figure}[h]
  \centerline{\resizebox{6cm}{!}{\includegraphics{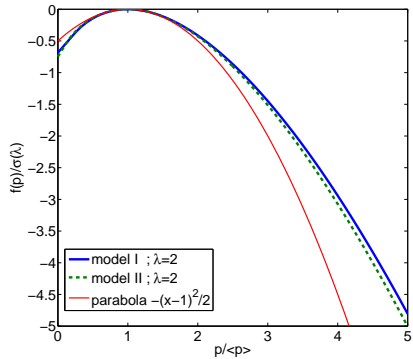}}}
\caption{Rescaled large deviation functions for $\la=2$ in both models.}\label{comparison_I_II}
\end{figure}

\begin{figure}[h]
  \centerline{\resizebox{6cm}{!}{\includegraphics{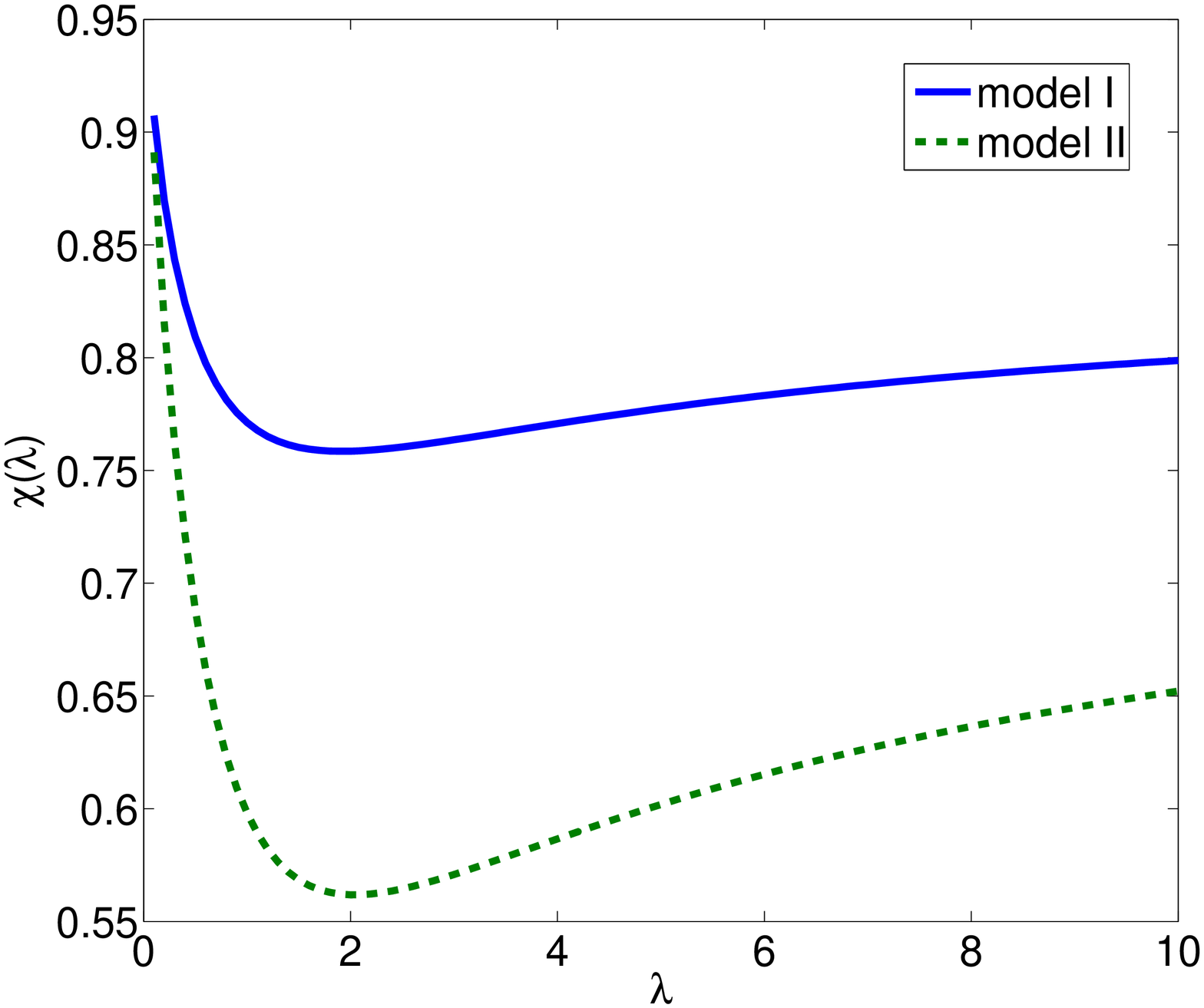}}}
\caption{Asymmetry parameter $\chi(\la)$. See text for details}\label{f3}
\end{figure}

We then  use  $\sig(\la)$ to rescale the ldfs so that the
curvatures are equal to 1, namely consider $f(p)/\sig(\la)$ as a function of $p/\lan
p\ran$. The results are plotted for  model I on figure \ref{res_I_rescaled_xy},
together with the reference parabola $-\demi (p/\lan p\ran-1)^2$. Corresponding
 curves for model II are very similar, as can be seen on
figure \ref{comparison_I_II}, where the ldfs for both systems with $\la=2$
(maximum discrepancy of parameters $\chi$, see later) are
simultaneously plotted.
We remark that the ldf are rather different from a parabola, and
display a marked tilt counterclockwise. Interestingly enough, the
magnitude of the tilt is not a monotonous function of $\la$. This can
be simply caught by inspecting $\chi(\la)=g'''(0)g'(0)/{g''}^2(0)$, the third coefficient of the Taylor
expansion of the scaled ldf
near the maximum: $f(p)/\sig(\la)+\demi(p/\lan p\ran-1)^2\simeq
\frac{1}{6}\chi (p/\lan p\ran-1)^3$. As shown on
figure \ref{f3}, $\chi(\la)$ has a non-monotonous
behaviour with respect to $\la$, and mainly dictates the location of  the
asymptotic tail. 
Note that this parameter is positive, in accordance
with the counterclockwise tilt already mentioned. The physical origin
of these features is rather complicated. The tilt is mainly associated
to the temporal correlations of the process, which are revealed in the
statistics of rare events only. The probability of an
extra injection of energy is more likely than the opposite, because of
a kind of positive feedback : if for a while, an extra amount of
energy has been already transfered into the system, the typical
density of domain walls near the boundary experiences a positive
fluctuation; as a result,  subsequent incoming domain walls are more
likely to eventually annihilate with an
alter ego, instead of returning to the boundary (conversely a negative fluctuation is hard to maintain due to
a growing shortage of domain walls to react with, see \cite{farago1} for similar arguments). 

\begin{figure}[h]
  \centerline{\resizebox{12cm}{!}{\includegraphics{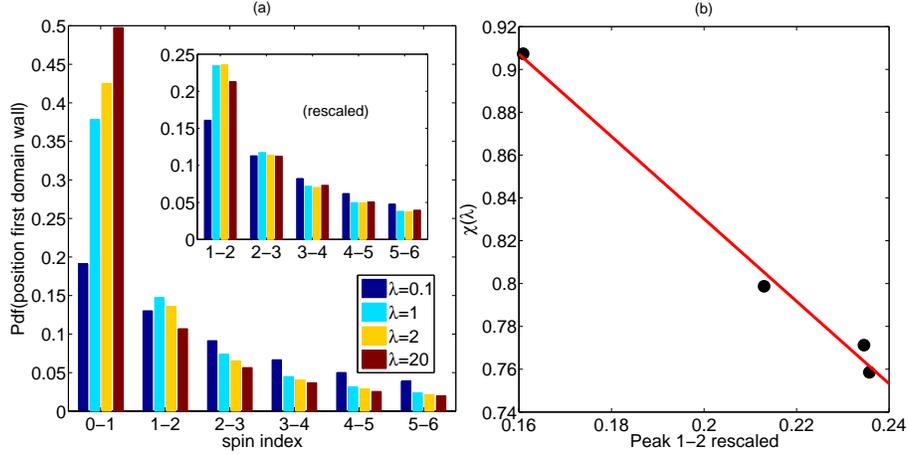}}}
\caption{(a) Repartition function of the domain wall nearest to the spin
  $0$ for model I. The inset shows the distribution function
  restricted (and rescaled in accordance) to the bulk domain
  walls. (b) Correlation plot between the height of the first peak
  (position 1-2) of the rescaled repartition function and
  $\chi(\lambda)$. The line is a linear fit.}\label{barplot}
\end{figure}

That this positive
feedback has a minimum is related to the structure of the stationary
state. On figure \ref{barplot} is plotted the repartition function of
the domain wall \underline{nearest to $s_0$} in the stationary state for model I (we restrict
the discussion to model I for sake of simplicity) and various values
of $\la$ ---these pdfs were obtained numerically although the exact
computation is a priori tractable. It is seen that the probability for the first domain wall to
be between spin 1 and spin 2 (that is, next to the first possible
domain wall, which so to speak does not belong  to the bulk as it is
directly affected by the boundary flips) is
maximum for $\la\sim 1-2$ (and is maximum with respect to the subsequent
locations). This property is also true if one disregards the 
domain walls located in contact with the spin $s_0$ and restricts
oneself to ``inner'' domain walls (see inset of fig \ref{barplot}).
The correspondence between $\chi(\la)$ and the rescaled height of the
peak 1-2 can be made quantitative by a correlation plot
(fig. \ref{barplot} (b)), where a linear relation between the
two quantities can be approximately drawn.

Thus, a density fluctuation is typically localized
very near the boundary especially for $\la\sim 2$, and this
proximity works against the correlations and the positive feedback effect explained above, as after just one move
these domain walls can disappear through a negative energy injection
event; this explains roughly the minimum of $\chi(\la)$ near $\la\sim 2$.
 Finally one explains also the less pronounced effect for
model II by the fact that the two halves of the system see each other
in that case and domain walls from one side can pervade  the other: this permeability
probably reduces the correlations between  density fluctuations and
 injection,  thereby decreasing the asymmetry parameter.

\medskip

\begin{figure}[h]
  \centerline{\resizebox{6cm}{!}{\includegraphics{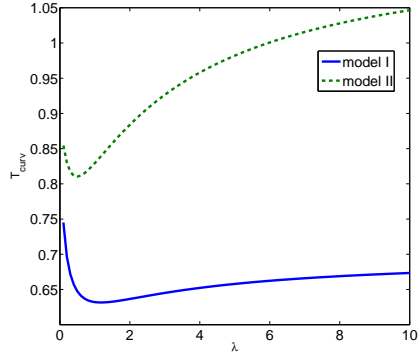}}}
\caption{Nonequilibrium temperature
  $T_\text{curv}(\la)=g''(0)/2g'(0)$. See text for details.}\label{Tcurv}
\end{figure}
Interestingly enough, this parameter $\chi$ has a similar behaviour as
the nonequilibrium ``temperature'' one can define for such dissipative
NESS from $f(p)$ \cite{farago2,affm} : $T_\text{curv}=g''(0)/2g'(0)$. Figure
\ref{Tcurv} shows this quantity for both models, and it is clear that
this characteristic energy constructed from the fluctuations of $\Pi$
is sensitive to both the temporal correlations of the process (as
$\chi$) and the averages quantities like the mean injected power. As
such, it bears a rather complicated physical content, in contrast with
an ordinary temperature concept.

\bigskip

Before to conclude, it is worth mentioning a mathematical subtlety of
model II: the naive $\la\rightarrow\infty$ limit yields
\begin{align}
  g(\al)\simeq \frac{2}{\pi}\int_0^\infty du\ \log\left|1+\frac{2(e^{2\al}-1)(1+i(\psi-1)^2/8u\psi)}{\psi+1}\right|^2\label{faux}
\end{align}
whence one would have
\begin{align}
  g'(0)=\lan p\ran&=\frac{16}{\pi}\text{Re}\left(\int_0^\infty du
  \frac{1+i(\psi-1)^2/8u\psi}{\psi+1}\right)\\
&=1
\end{align}
But this is not the correct result ($\lan p\ran=2$), and it is easy to
see why : the exact formula for finite $\la$ is, provided
$\sqrt{-u^2+iu}=R(u)+iI(u)$ :
\begin{align}
  g'(0)&=\frac{8\la}{\pi}\int_0^\infty du \ \frac{\la/2+R+I(\psi-1)^2/8u\psi}{(\psi+1)([\la/2+R]^2+I^2)}\\
  R(u)&=\frac{1}{\sqrt{2}} \sqrt{-u^2+\sqrt{u^4+u^2}}\\
  I(u)&=\frac{1}{\sqrt{2}} \sqrt{u^2+\sqrt{u^4+u^2}}
\end{align}
The missing part comes from the third term of the numerator : $I\sim
u$ for large $u$, hence the associated integral is of order $1/\la$
and not $1/\la^2$ as naively expected. This problem is ultimately due to the fact
that an isolated pole exists for model II at the right hand side of equation (\ref{superultimate})
 (see section \ref{calcul}), an analytical property extremely sensitive to the
form of the approximants: some sub-dominants terms are of primarily
importance to keep the analytical structure untouched and makes allowance for
the asymptotic locations of the singularities. 

\section{Conclusion}

In this paper, we were able to compute exactly by fermionic techniques
 large deviation functions of the injected power for two 1D models of
classical spins driven by a $T=0$ Glauber dynamics in the bulk and a
Poissonian flip on the boundary. The results highlight the influence
of the flipping rate  on the low-frequency fluctuation properties of
the injection; we showed in particular that the third derivative of $f$ is a sensible
measure of the  interplay between on one hand density
fluctuations and on the other hand
the ability for the external operator to transfer efficiently 
energy inside the system. By the way, we found and explained why  this third
derivative (or equivalently the third cumulant of the time-integrated injected energy)
is always positive, a property surprisingly found also in conservative
diffusive systems for the integrated current (see formula (4) in
\cite{bodineauderrida}); this similar behaviour is a bit surprising, as inner dynamics of dissipative and conservative
systems are by nature  much different. 

\section*{Acknowledgements}

We are much indebted to S. Auma\^\i tre, F. Cornu, S. Fauve and  F. van Wijland for
fruitful discussions. This work was supported by the
ANR project JCJC-CHEF.

\section{Appendix}
We give some details on the diagonalization of a Hamiltonian of the
type:
\begin{align}\label{Happ}
  H&=\sum_{n,m}\left[c^\dag_nA_{nm}c_m+\demi c_n^\dag B_{nm}c_m^\dag+\demi c_n
   D_{n,m}c_m\right]\end{align}
where $A$ is symmetric real and $B$ and $D$ antisymmetric real.
We postulate the transformation
\begin{align}\label{ccapp}
  c^\dag&=V\xi^\dag+U\xi\\
c&=U^*\xi^\dag+V^*\xi
\end{align}
where $U$ and $V$ are $2N\times 2N$  matrices. The fermionic
structure is preserved if
\begin{align}
&  UU^{*T}+VV^{*T}=1\\
& UV^T+VU^T=0
\end{align}
It must be stressed that these equations are only $2N(2N+1)$
independent relations among $2(2N)^2$ coefficients. Thus $2N(2N-1)$
relations can still be imposed to the coefficients of $U$ and $V$.

Next, we can replace \myref{ccapp} into \myref{Happ} and extract four
different contributions : a contribution made with terms $\xi\xi$
($N(2N-1)$ terms) and
$\xi^\dag\xi^\dag$ ($N(2N-1)$ terms) we want to get rid of :
\begin{align}
  \demi(\pmb{\xi^\dag})^T\left[V^TAU^*-U^{*T}AV+ V^TBV+
  U^{*T}DU^*\right](\pmb{\xi^\dag})\\
  \demi(\pmb{\xi})^T\left[U^TAV^*-V^{*T}AU+ U^TBU+
  V^{*T}DV^*\right](\pmb{\xi})
\end{align}
this
gives precisely the $2N(2N-1)$ remaining relations on $U$ and $V$ (the
matrices are antisymmetric);
thirdly a contribution $\xi^\dag \xi$ 
\begin{align}
  (\pmb{\xi^\dag})^T M (\pmb{\xi})&\sur{=}{\text{def}}  (\pmb{\xi^\dag})^T\left[V^TAV^*-U^{T}AU^*+\demi( V^TBU-U^TBV)+\demi(U^{*T}DV^*-V^{*T}DU^*)
\right](\pmb{\xi})
\end{align}
Our goal is to diagonalize this matrix. If this is possible, the
diagonalization is completed using $M=Q^{-1}\De_MQ$ and
$\zeta=Q\xi,\zeta^\dag=Q^{-1 T}\xi^\dag$. 
But this is  apparently not always
possible, as the inspection of $M$ clearly shows.
  Two special situations allow however a diagonalization: if $B=-D$,
  $M$ is hermitian and therefore diagonalizable ; if $U$ and $V$ can be
  found real, $M$ is symmetric and also diagonalizable. But the last
  case we cannot immediately recognize. In the following we
  \textit{assume} that a  diagonalization of $M$ can be completed.

There is a fourth term in the decomposition, which is a constant
arisen from the commutation $\xi\xi^\dag\rightarrow\xi^\dag\xi$. This
constant reads
\begin{align}
  \text{Tr}\left[U^TAU^*+\demi U^TBV+\demi V^{*T}DU^*\right]=-\demi\text{Tr}(M)+\demi\text{Tr}(A)
\end{align}
where relations among $U$ and $V$ were used.
As a result, the Hamiltonian can be cast in the following form :
\begin{align}\label{Hdiag}
  H=\sum_q\La_q\left(\xi_q^\dag\xi_q-\demi\right)+\demi\text{Tr}(A)
\end{align}
where the $\La_q$ are the eigenvalues of $M$. This formula differs slightly from those presented in \cite{henkel}
where the Tr$(A)$ term is not written. It is interesting to see that
the precise sign of $\La_q$ is irrelevant, as far as the spectrum of
$H$ is concerned. The eigenvalues of $H$ are actually given by
\begin{align}
  \demi\left(\sum_q \La_q\eps_q+\text{Tr}(A)\right)
\end{align}
where the $\eps_q$ are $\pm 1$.

\medskip

The preceding development essentially yields the conclusion embodied
by the equation \myref{Hdiag}, provided that the diagonalization is
possible. To study the diagonalization itself, it is particularly
clever to follow the route described in \cite{henkel}, where in
\myref{ccapp}, the roles of $c$ and $\xi$ are reversed. Starting from
\myref{Hdiag} and using $[H,\xi_q]=\La_q\xi_q$, one gets the following
result: the spectrum of the matrix
\begin{align}
  M_0=\left(
  \begin{array}{cc}
    A&B\\D&-A
  \end{array}\right)
\end{align}
is made with the eigenvalues of $M$. Actually it is exactly the
spectrum of $M$ plus its symmetric with respect to zero : multiplying $\mu\text{Id}-M_0$ by
\begin{align}
  \left(
  \begin{array}{cc}
    (\mu-A)^{-1}&0\\ (\mu+A)^{-1}D(\mu-A)^{-1} & (\mu+A)^{-1}
  \end{array}\right)
\end{align}
it is easily seen that the characteristic polynomial
$\chi_{M_0}(\mu)=\det(\mu\text{Id}-M_0)$ can be recast into
\begin{align}\label{app47}
\chi_{M_0}(\mu)=\chi_A(\mu)\chi_A(-\mu)(-1)^{\text{dim}(A)}\det(1-B(\mu+A)^{-1}D(\mu-A)^{-1})
\end{align}
Next, using the relation $\det(1+AB)=\det(1+BA)$ valid whatever $A$
and $B$, and $\det(A)=\det(A^T)$, it is readily seen that
$\chi_{M_0}(\mu)=\chi_{M_0}(-\mu)$. Thus, the spectrum of $H$ is just
duplicated in $M_0$ according to $\text{Sp}(M_0)=\{\text{Sp}(H),-\text{Sp}(H)\}$

\medskip

The question of the diagonalization of $H$ is closely related to the
diagonalization of $M_0$. The case $D=-B$ is explicit in this
representation since in that case $M_0$ is symmetric and thus
diagonalizable with real eigenvalues. Actually, as the non
diagonalization of a matrix is an accident rather than the rule, we
can consider these cases as exceptional.

\end{document}